\documentclass[a4paper]{spie}  
\usepackage[]{graphicx}
\usepackage{amssymb,amsmath,psfrag,url}
\usepackage{verbatim}

\newcommand{\Bolivarallee}{Boliva\hspace{-0.1mm}r\hspace{0.15mm}a\hspace{-0.1mm}llee}

\newcommand{\Takustrasse}{Taku\hspace{0.25mm}s\hspace{-0.1mm}tra{\ss}e}

\newcommand{\Field}[1]{{\bf{#1}}}
\newcommand{\Tensor}[1]{{\bf{#1}}}

\newcommand{\curl}{\mathbf{curl}\;}

\title{Fast Simulation Method\\ for Parameter Reconstruction\\ in Optical Metrology}

\author{
Sven~Burger,\supit{\,ab}
Lin~Zschiedrich,\supit{\,b}
Jan~Pomplun,\supit{\,b}\\
Frank~Schmidt,\supit{\,ab}
Bernd~Bodermann\supit{\,c}
\skiplinehalf
\supit{a}
Zuse Institute Berlin\,(ZIB),
\Takustrasse~7,
D\,--\,14\,195 Berlin,
Germany
\smallskip\\
\supit{b}
JCMwave GmbH,
\Bolivarallee~22, 
D\,--\,14\,050 Berlin,
Germany
\smallskip\\
\supit{c}
Physikalisch-Technische Bundesanstalt Braunschweig (PTB), \\ 
Bundesallee~100, D\,--\,38\,116 Braunschweig, Germany
}
\authorinfo{
Corresponding author: S. Burger\\
URL: http://www.zib.de\\
URL: http://www.jcmwave.com\\
Email: burger@zib.de
}

\begin{document}
\maketitle

\noindent
This paper will be published in Proc.~SPIE Vol. {\bf 8681}
(2013) 868119, ({\it Metrology, Inspection, and 
Process Control for Microlithography XXVII}, DOI: 10.1117/12.2011154), 
and is made available 
as an electronic preprint with permission of SPIE. 
One print or electronic copy may be made for personal use only. 
Systematic or multiple reproduction, distribution to multiple 
locations via electronic or other means, duplication of any 
material in this paper for a fee or for commercial purposes, 
or modification of the content of the paper are prohibited.

\begin{abstract}
A method for automatic computation of parameter derivatives 
of numerically computed light scattering signals is demonstrated. 
The finite-element based method is validated in a numerical convergence study, and it 
is applied to investigate the sensitivity of a scatterometric setup 
with respect to geometrical parameters of the scattering target.  
The method can significantly improve numerical performance of design optimization, 
parameter reconstruction, sensitivity analysis, and other applications. 
\end{abstract}


\section{Introduction}
In optical metrology of nanostructures rigorous (i.e., accurate) simulation of light propagation is an essential 
component~\cite{Pang2012aot,Lai2012aot}. 
A challenge consists in reducing computation times for simulation results matching predefined accuracy requirements. 
This is especially important when real-world structures of complex geometry are considered. 

\begin{figure}[b]
\begin{center}
  \includegraphics[width=.7\textwidth]{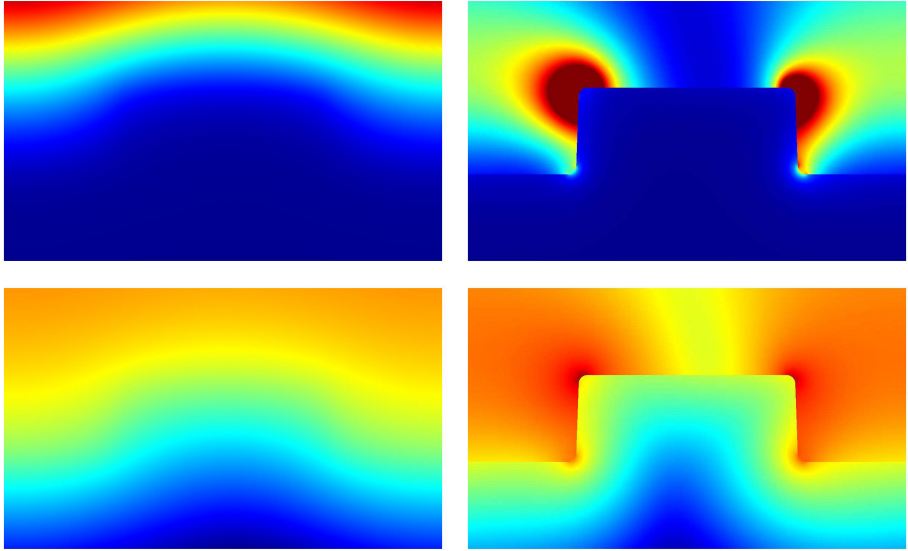}
\caption{Electric field intensity distribution in pseudo-color representation (red: high intensity, blue: low intensity). 
         Top: linear color scale, bottom: logarithmic color scale. Left: S-polarized light, right: P-polarized light. 
}
\label{fields}
\end{center}
\end{figure}

We present a fast, finite-element based method to address such computation challenges. 
In this contribution we especially focus on finite-element based computation of 
derivatives of the propagating light fields (and of derived quantities like transmission or reflection intensities) 
with respect to geometrical parameters of the scattering target. 
As practical example we present a sensitivity analysis for patterns on a scatterometry 
reference standard: 
dependence of the scatterometric signal on geometry parameters 
(CDs, sidewall-angles, corner-rounding) is evaluated in various parameter regimes.

This paper is structured as follows: 
The background of our model is presented in Section~\ref{section_background}, 
the numerical method is described in Section~\ref{section_numerical}, 
convergence results are reported in Section~\ref{section_convergence}, 
and results of a sensitivity analysis of scatterometric signals from a pattern proposed as sample on a scatterometric 
standard is reported in Section~\ref{section_sensitivity}.

\section{Background / Model}
\label{section_background}
Light scattering off nanoscopic structures on scatterometry samples is modeled by 
the linear Maxwell's equations in frequency domain~\cite{Pomplun2007pssb,Burger2012springer}. 
From these a single equation for the electric field $\Field{E}$ can be derived:
\begin{equation}
  \curl\Tensor{\mu}^{-1}\curl \Field{E}-\omega^{2}\Tensor{\epsilon}\Field{E}=i\omega\Field{J}.
  \label{eq:mwE}
\end{equation}
where  $\Tensor{\epsilon}$ and $\Tensor{\mu}$ are the permittivity and permeability tensor, $\omega$ is 
the time-harmonic frequency of the electromagnetic field, and the  
electric current $\Field{J}$ is source of an electromagnetic field. 
The domain of interest is separated into an infinite 
exterior $\Omega_{\mathrm{ext}}$ which hosts the given incident field and the scattered field, 
and an interior $\Omega_{\mathrm{int}}$ where the total field is computed. 
Electromagnetic waves incident from the exterior to the interior at the boundaries between both domains
are added to the right hand side of Eq.~\eqref{eq:mwE}. 
For numerical simulations the infinite exterior is treated using transparent 
boundary conditions (using the perfectly matched layer method, PML).

Transforming Eq.~\eqref{eq:mwE} into weak formulation and discretizing it using 
finite elements yields a matrix equation:
\begin{equation}
  A \Field{E}_h = f
  \label{eq:matrixE}
\end{equation}
where $A$ is a sparse matrix, $f$ contains the source terms, 
and $\Field{E}_h$ is the expansion of the electric field in a finite-dimensional FEM basis. 

Inversion of $A$ and multiplication with the right hand side gives the  solution $\Field{E}_h$:
\begin{equation}
  \Field{E}_h = A^{-1}f
  \label{eq:invmatrixE}
\end{equation}

Note that solutions corresponding to different sources incident on the same pattern can be obtained from the 
same inverted system matrix, given that $A$ does not depend on the sources. 
E.g., when $f_1$ and $f_2$ correspond to incident light of two different polarizations, the corresponding near fields 
$\Field{E}_{h,1}$ and $\Field{E}_{h,2}$   can be obtained from the same inverted system matrix:

\begin{eqnarray}
  \Field{E}_{h,1}& =& A^{-1}f_1\\
  \Field{E}_{h,2}& = &A^{-1}f_2
  \label{eq:invmatrixE2}
\end{eqnarray}

Inversion of the system matrix (i.e., computation of $A^{-1}$) 
typically is the computationally most costly step, therefore {\it re-using} the same 
inverted matrix $A^{-1}$ for $N$ sources reduces the computational costs approximately by a factor of $N^{-1}$, 
in a simulation setting where $N$ independent source terms are present.

In optimization problems, reconstruction problems and sensitivity studies, often an accurate measure of the 
partial derivative of the near field with respect to project parameters $p_i$ 
(e.g., geometry parameters, source parameters, material parameters), $\partial p_i\Field{E}_{h}$, is required. 
As is well known,  it is straight-forward in the finite-element context to compute these quantities 
by again {\it re-using} the inverted system matrix:
\begin{equation}
  \partial p_i\Field{E}_{h} = A^{-1}[\partial p_i f -(\partial p_i A )\Field{E}_{h}] 
  \label{eq:partial1}
\end{equation}
Also higher-order derivatives $\partial^N p_i\Field{E}_{h}$ can be computed, e.g., 
\begin{equation}
  \partial^2 p_i\Field{E}_{h} = 
A^{-1}[(\partial^2 p_i f) -(\partial^2 p_i A )\Field{E}_{h} - 2(\partial p_i A)(\partial p_i \Field{E}_{h})] 
  \label{eq:partial2}
\end{equation}
Here, $\partial^N p_i A$ is the $N$th derivative of $A$ with respect to parameter $p_i$, and 
$\partial^N p_i f$ is the $N$th derivative of source term $f$ with respect to parameter $p_i$.

\begin{figure}[t]
\begin{center}
\psfrag{R}{\sffamily $R$}
\psfrag{alpha}{\sffamily $\alpha$}
\psfrag{ha}{\sffamily $h$}
\psfrag{line}{\sffamily line}
\psfrag{superspace}{\sffamily superspace}
\psfrag{substrate}{\sffamily substrate}
\psfrag{px}{\sffamily $p$}
\psfrag{cd}{\sffamily $CD$}
  \includegraphics[width=.5\textwidth]{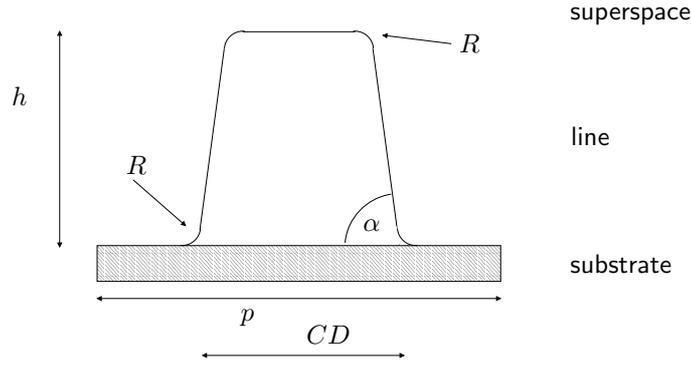}
  \caption{
Schematics of the geometry of the investigated scatterometric target (unit cell of a 1D-periodic grating). 
Free parameters of the model are the critical dimension (CD, width at $h/2$), the height $h$, pitch $p$, sidewall 
angle $\alpha$, corner rounding radius $R$. 
}
\label{schematics_emrp_1}
\end{center}
\end{figure}

\section{Numerical method}
\label{section_numerical}
For rigorous simulations of the scattered light field we use the 
finite-element (FEM) Maxwell solver JCMsuite.
This solver incorporates higher-order edge-elements, self-adaptive meshing, 
and fast solution algorithms for solving time-harmonic Maxwell's equations. 
Also, automatic computation of first- and higher-order parameter derivatives is 
implemented in the software. 
Previously the solver has, e.g.,  been used in scatterometric investigations   
of EUV line masks (1D-periodic patterns), contact hole masks 
(2D-periodic patterns) and more complicated 3D patterns~\cite{Scholze2007a,Scholze2008a,Burger2011eom1,Burger2011pm1,Kato2012a}.
Convergence studies in these investigations demonstrate that highly  accurate, rigorous 
results can be attained even for the relatively large 3D computational domains which are typically present 
in 3D EUV setups. 

The workflow for the simulations is as follows:
a scripting language (Matlab) automatically iterates the input parameter sets 
(physical parameters like geometrical dimensions and numerical parameters like mesh refinement).
For each set, a triangular 2D mesh is created automatically by the built-in mesh generator. 
Then, the solver is 
started for computing the electromagnetic near field and its parameter derivatives, 
postprocessing is performed to extract, e.g., diffraction order efficiencies and their parameter derivatives, 
and results are evaluated and saved. 

\begin{table}[h]
\begin{center}
\begin{tabular}{|l|l|}
\hline
dimension & 1D \\ \hline
material & Si  \\ \hline
pitch & 100\,nm\\ \hline
CD & 50\,nm\\\hline
$h$& 20\,nm \\ \hline
$\alpha$& 88\,deg \\ \hline
$R$ & 2\,nm\\ \hline
$\lambda$& 193\,nm \\ \hline
$\theta$& 30\,deg \\ \hline
$\phi$& 0\,deg \\ \hline
\end{tabular}
\caption{Parameter settings for the scatterometry standard simulations (compare Fig.~\ref{schematics_emrp_1}).
Line height $h$, sidewall angle $\alpha$, corner rounding radius $R$, illumination vacuum wavelength $\lambda_0$, 
illumination inclination and rotation angle, $\theta$ and $\phi$.
Parameter settings for the scatterometry standard simulations (compare Fig.~\ref{schematics_emrp_1})
}
\label{table_specs_2}
\end{center}
\end{table}

\begin{figure}[t]
\begin{center}
  \includegraphics[width=.7\textwidth]{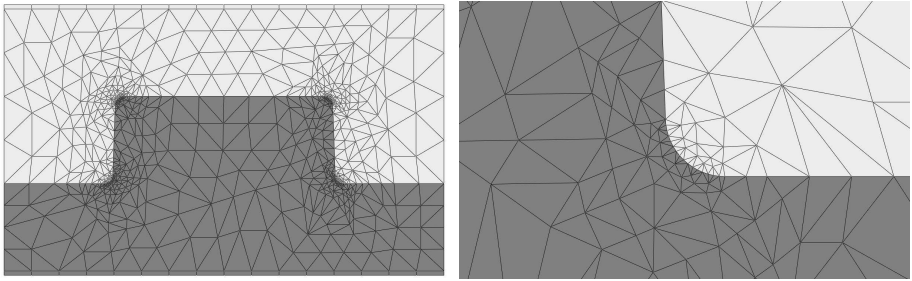}
\caption{Finite-element mesh for spatial discretization of the geometry. Left: full geometry, right: detail at a rounded 
corner. 
}
\label{mesh}
\end{center}
\end{figure}

Numerical settings which yield highly accurate results for the setup of interest in the presented investigations
are identified in a convergence study 
(Section~\ref{section_convergence}). 
As numerical settings for the solver in the subsequent Section~\ref{section_sensitivity} on a sensitivity study, 
finite elements of third-order polynomial degree, and adaptive, error-estimator controlled meshing of the 
geometry in the computational domain and of transparent boundaries are chosen. 
This setting yields discrete problems with few ten thousands of unknowns (e.g., 30,000 unknowns), 
and few seconds (e.g., 4\,sec) of computation time
per computation (for computation of reflectivities and their parameter derivatives, for two polarizations, and 
for a specific physical setting).
The FEM software solves these problems by direct LU factorization on a standard desktop computer. 
Figure~\ref{fields} shows a graphical representation of a typical near-field intensity distribution. Please note that 
(as expected for this angle of incidence) the S-polarized incident wave leads to a smooth intensity distribution 
while the P-polarized incident wave leads to a highly discontinuous intensity distribution. 

\section{Model validation}
\label{section_convergence}

\begin{figure}[t]
\begin{center}
\begin{minipage}[c]{.45\textwidth}
  \psfrag{Relative error}{\sffamily Relative error}  
  \psfrag{p}{\sffamily $p$}
  \psfrag{Rs}{\sffamily $I_0$}
  \psfrag{dw Rs}{\sffamily $\partial I_0/\partial_{CD}$}
  \psfrag{dh Rs}{\sffamily $\partial I_0/\partial_{h}$}
  \psfrag{dswa Rs}{\sffamily $\partial I_0/\partial_{\alpha}$}
  \includegraphics[width=.97\textwidth]{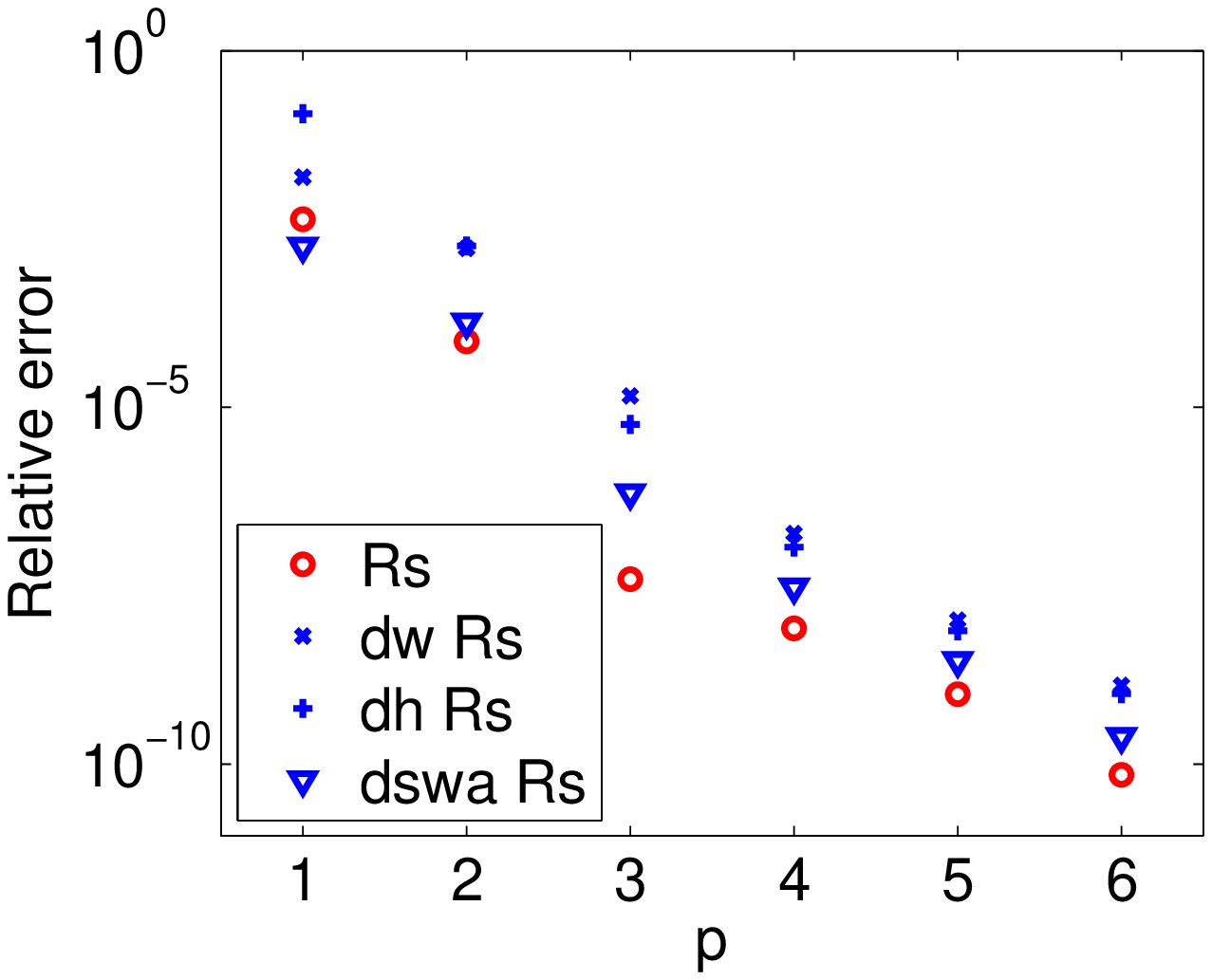}
 \end{minipage}
\begin{minipage}[c]{.45\textwidth}
  \psfrag{Relative error}{\sffamily Relative error}  
  \psfrag{p}{\sffamily $p$}

  \psfrag{Rp}{\sffamily $I_0$}
  \psfrag{dw Rp}{\sffamily $\partial I_0/\partial_{CD}$}
  \psfrag{dh Rp}{\sffamily $\partial I_0/\partial_{h}$}
  \psfrag{dswa Rp}{\sffamily $\partial I_0/\partial_{\alpha}$}
  \includegraphics[width=.97\textwidth]{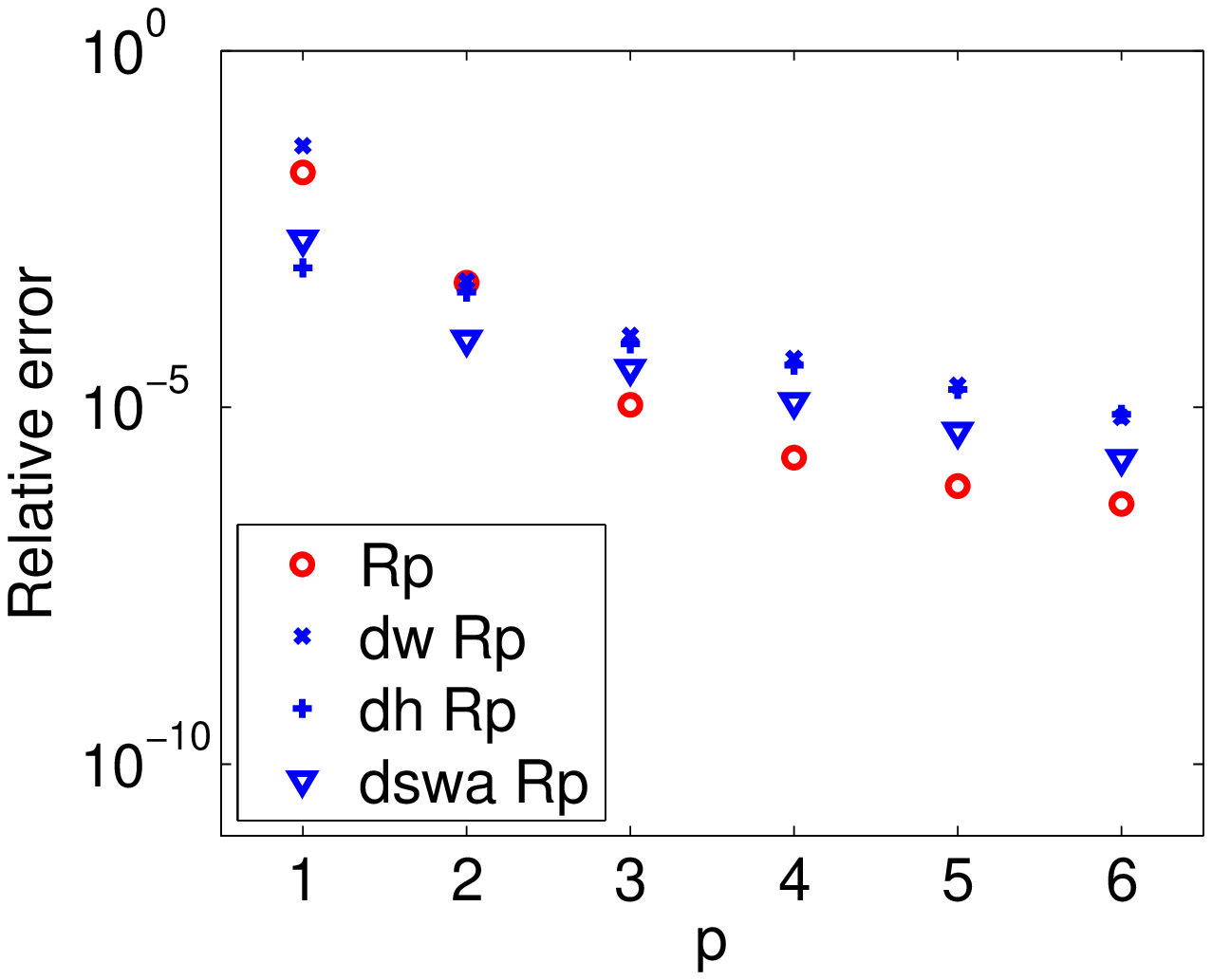}
\end{minipage}
  \caption{
Dependence of the relative error of the reflectivity and its derivatives with respect to geometry parameters 
on numerical parameter $p$. 
{\it Left:} S-polarized incident light, 
{\it Right:} P-polarized incident light.
}
\label{figure_convergence}
\end{center}
\end{figure}

In order to validate our model we perform a convergence study where we investigate how the computed quantities 
and their derivatives with respect to geometry parameters depend on the chosen numerical parameters. 
We investigate a geometry which could be used as part of a scatterometric standard~\cite{Bodermann2012op}. 
The investigated pattern is a 1D-periodic line gratings 
etched into silicon (Si), with specific pitch (periodicity) and center line-width (CD) 
Figure~\ref{schematics_emrp_1} 
shows a schematics of the 2D setup for this test case. 
Table~\ref{table_specs_2}
shows parameter values of the project setup.
Figure~\ref{mesh} shows a graphical representation of a 2D mesh.

The pattern is illuminated from the superspace at oblique incidence with S- and P-polarized, monochromatic 
plane waves. 
The quantity of interest in this case is the intensity of light in the zero'th reflected diffraction order, 
$I_0$ ($I_0\sim  |\Field{E}|^2$, cf., Eq.~\ref{eq:mwE}), and 
it's derivatives with respect to line width, height and sidewall angle, 
$\partial I_0/\partial_{CD}$, 
$\partial I_0/\partial_{h}$, 
$\partial I_0/\partial_{\alpha}$, 
as function of varied geometry parameters. 
Please note that here, we normalize $I_0$ with the intensity of the incoming light field, i.e., $I_0$ is 
a dimensionless quantity.
This numerical study is restricted to evaluation of intensities of the unpolarized light field, $I_0$, 
however, as the derivatives of the vectorial electric field amplitudes are computed ($\partial p_i\Field{E}_{h}$,
cf., Eq.~\ref{eq:partial1}), also other quantities (sensitivities of all entries in the M\"uller matrix) 
are accessible without extra computational costs.
This numerical study is also restricted to 1D-periodic patterns (i.e., 2D computational domains), 
however, the method can also be applied (and is implemented in the software) for 3D setups and/or isolated 
computational domains (i.e., non-periodic setups).

Numerical errors as present in any numerical method for solving Maxwell's equations
depend on the actual numerical settings. The two main numerical degrees of freedom for the finite-element method are 
the spatial discretization (mesh refinement) and the choice of ansatz functions which are used to approximate 
the fields on the spatial discretization mesh. The ansatz functions are typically defined by their polynomial degree $p$
(when ansatz functions with a higher degree are chosen, this results in a larger basis for approximating the 
solution, and -- more importantly -- in higher approximation quality~\cite{Pomplun2007pssb}). 
Figure~\ref{figure_convergence} shows how the numerical error of the reflection intensity and of its derivatives 
converges with finite element polynomial degree $p$. Relative errors are defined as normalized deviations from 
so-called quasi-exact results (results obtained at higher numerical discretization)~\cite{Burger2005bacus,Burger2008bacus}. 
As can be seen from this Figure, very high levels of accuracy are reached for both, the reflected intensities, and 
for their derivatives with respect to geometry parameters. We have also checked that computing these derivatives 
using numerical differentiation yields the same numerical values (however, with worse convergence properties, and at 
significantly higher numerical cost).

\section{Sensitivity study}
\label{section_sensitivity}

In order to demonstrate utility of the method we have performed several exemplary sensitivity studies. 
For given setups we investigate how the derivatives with respect to geometry parameters depend on specific 
physical parameter settings. This can be used to identify regimes where a scatterometric setup should work 
with higher sensitivity (yielding lower measurement uncertainties) than in other regimes. 

\begin{figure}[t]
\begin{center}
\begin{minipage}[c]{.45\textwidth}
  \psfrag{theta}{\sffamily $\theta$ [deg]}
  \psfrag{R}{\sffamily $I_0$}
  \psfrag{S-Pol}{\sffamily S-Pol}
  \psfrag{P-Pol}{\sffamily P-Pol}
  \includegraphics[width=.97\textwidth]{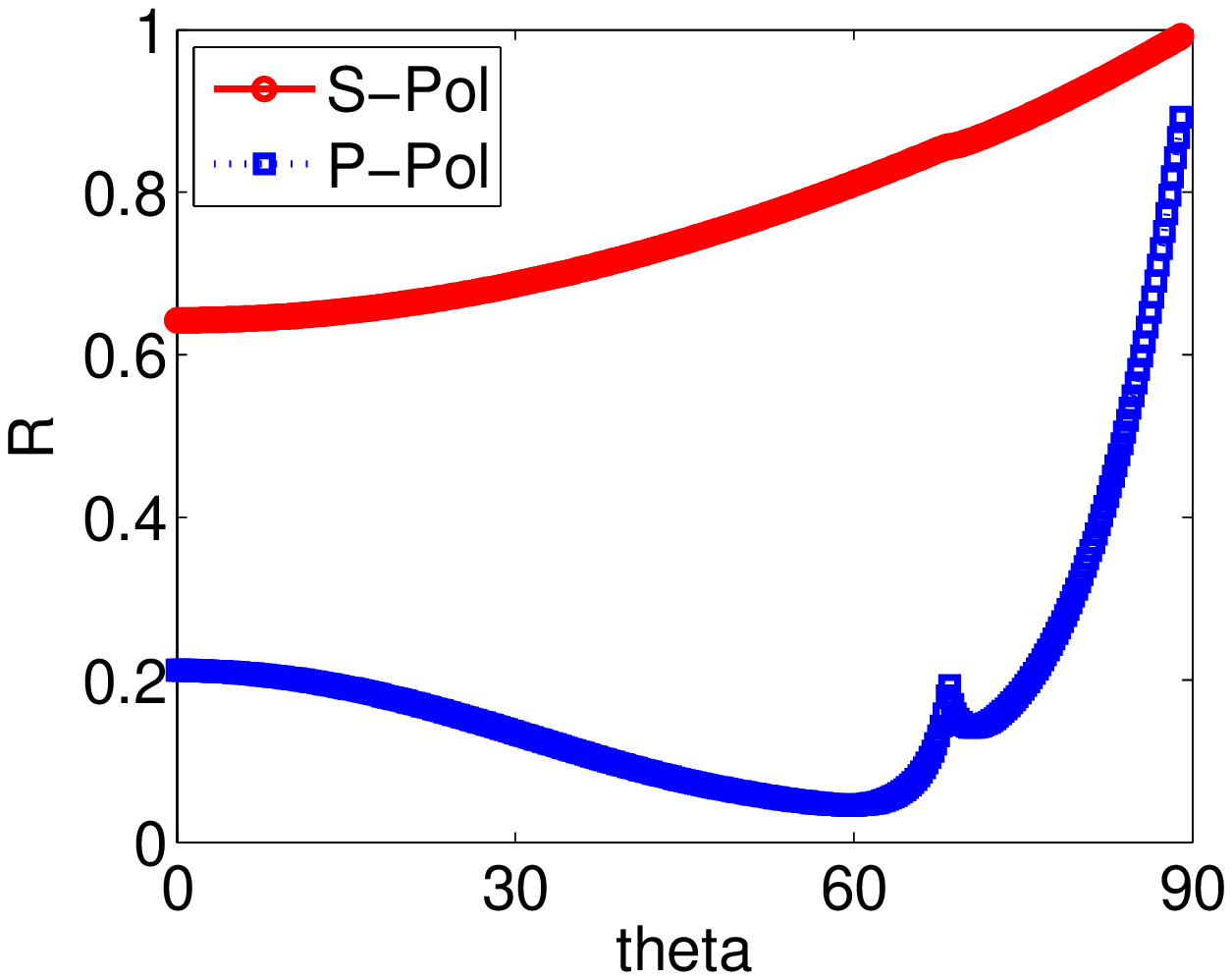}
 \end{minipage}
\begin{minipage}[c]{.45\textwidth}
  \psfrag{theta}{\sffamily $\theta$ [deg]}
  \psfrag{dp R}{\sffamily $\partial I_0/\partial_{p}$ [1/deg]}
  \psfrag{p, dw R p}{\footnotesize \sffamily P, $\partial I_0/\partial_{\tiny CD}$}
  \psfrag{p, dh R p}{\footnotesize \sffamily P, $\partial I_0/\partial_{h}$}
  \psfrag{p, dswa R p}{\footnotesize \sffamily P, $\partial I_0/\partial_{\alpha}$}
  \psfrag{s, dw R s}{\footnotesize \sffamily S, $\partial I_0/\partial_{\tiny CD}$}
  \psfrag{s, dh R s}{\footnotesize \sffamily S, $\partial I_0/\partial_{h}$}
  \psfrag{s, dswa R s}{\footnotesize \sffamily S, $\partial I_0/\partial_{\alpha}$}
  \includegraphics[width=.97\textwidth]{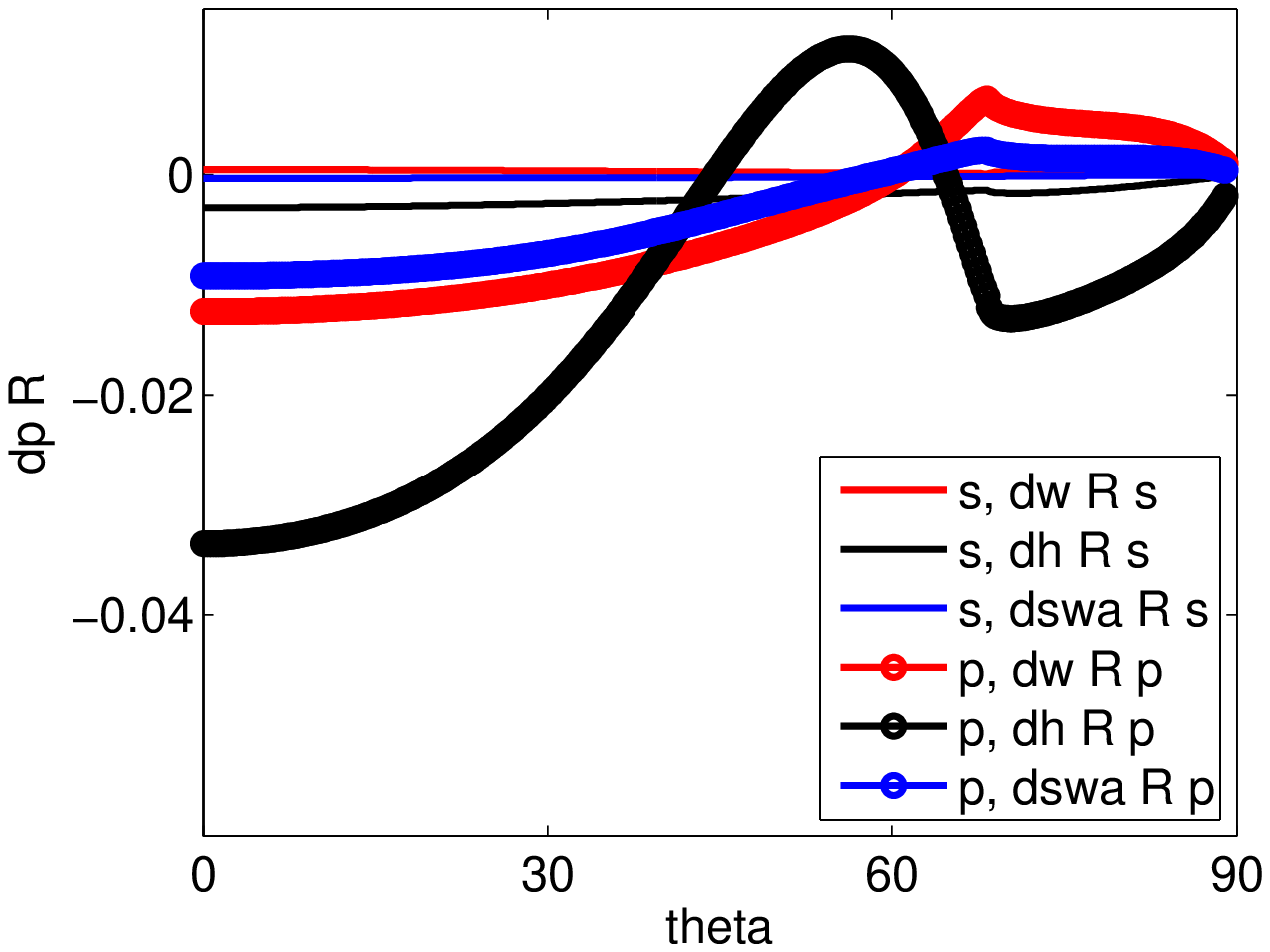}
\end{minipage}
  \caption{
{\it Left:} Dependence of the scatterometric signal $I_0$ on the azimuthal angle of incidence of the illuminating 
plane waves, for S- and P-polarization. 
{\it Right:} Dependence of the sensitivity with respect to parameter variations (CD, height, sidewall angle) on 
the angle of incidence. 
}
\label{figure_sensitivity_angle}
\end{center}
\end{figure}

Figure~\ref{figure_sensitivity_angle} (left) shows how the scatterometric signal (zero order reflection intensity) 
varies with azimuthal angle of incidence of the illuminating 
plane waves. As expected, S- and P-polarization show a different behavior. 
The right part of this Figure shows how the sensitivity with respect to parameter variations depends on this angle. 
From this Figure it can, e.g., be seen that in this case sensitivity is about an order of magnitude higher for incident P-polarized light, 
and that absolute values of sensitivity are highest for small angles $\theta$ (i.e., close to perpendicular incidence). 

\begin{figure}[t]
\begin{center}
\begin{minipage}[c]{.45\textwidth}
  \psfrag{h}{\sffamily $h$ [nm]}
  \psfrag{R}{\sffamily $I_0$}
  \psfrag{S-Pol}{\sffamily S-Pol}
  \psfrag{P-Pol}{\sffamily P-Pol}
  \includegraphics[width=.97\textwidth]{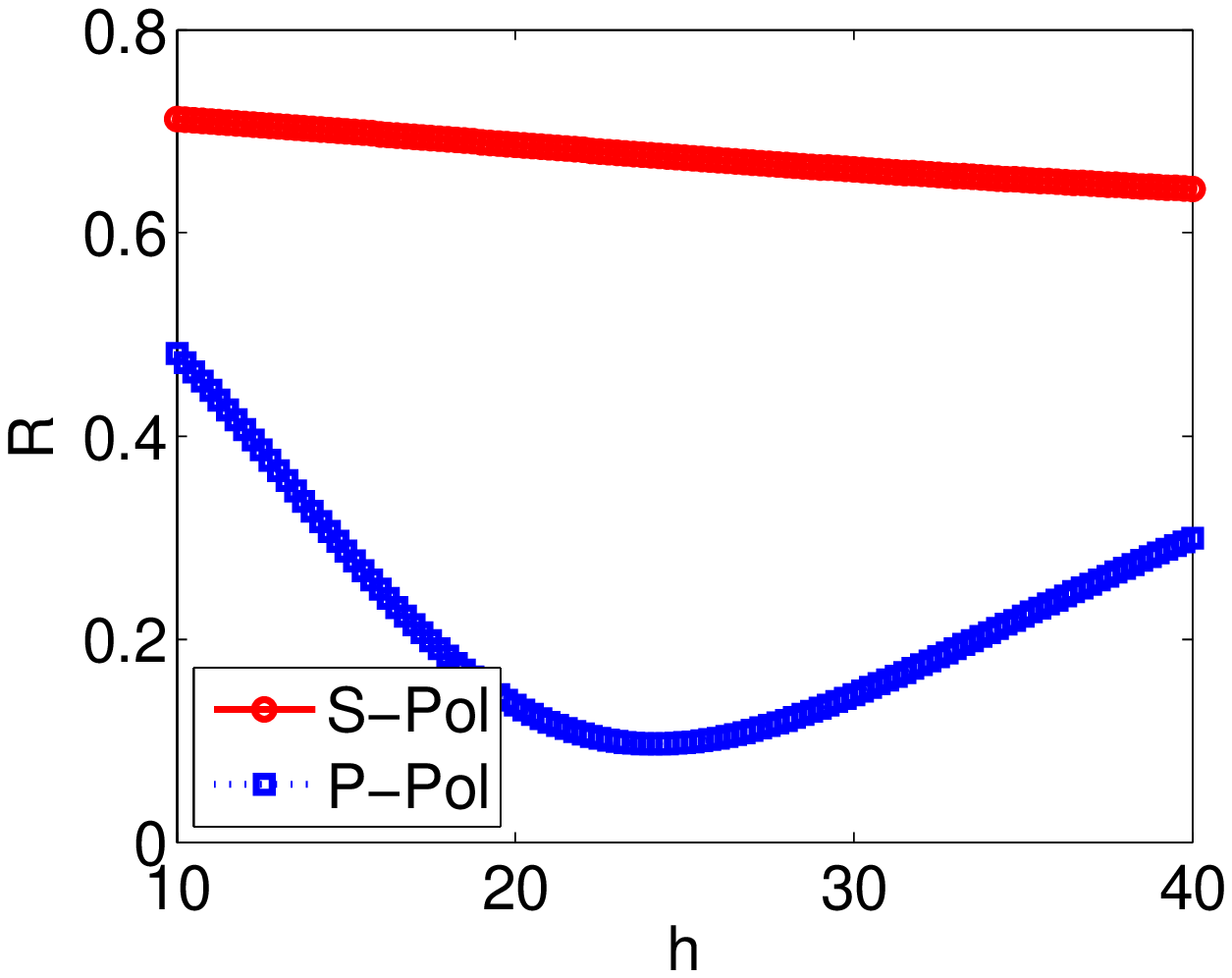}
 \end{minipage}
\begin{minipage}[c]{.45\textwidth}
  \psfrag{h}{\sffamily $h$ [nm]}
  \psfrag{dp R}{\sffamily $\partial I_0/\partial_{p}$  [1/nm]}
  \psfrag{p, dw R p}{\footnotesize \sffamily P, $\partial I_0/\partial_{\tiny CD}$}
  \psfrag{p, dh R p}{\footnotesize \sffamily P, $\partial I_0/\partial_{h}$}
  \psfrag{p, dswa R p}{\footnotesize \sffamily P, $\partial I_0/\partial_{\alpha}$}
  \psfrag{s, dw R s}{\footnotesize \sffamily S, $\partial I_0/\partial_{\tiny CD}$}
  \psfrag{s, dh R s}{\footnotesize \sffamily S, $\partial I_0/\partial_{h}$}
  \psfrag{s, dswa R s}{\footnotesize \sffamily S, $\partial I_0/\partial_{\alpha}$}
  \includegraphics[width=.97\textwidth]{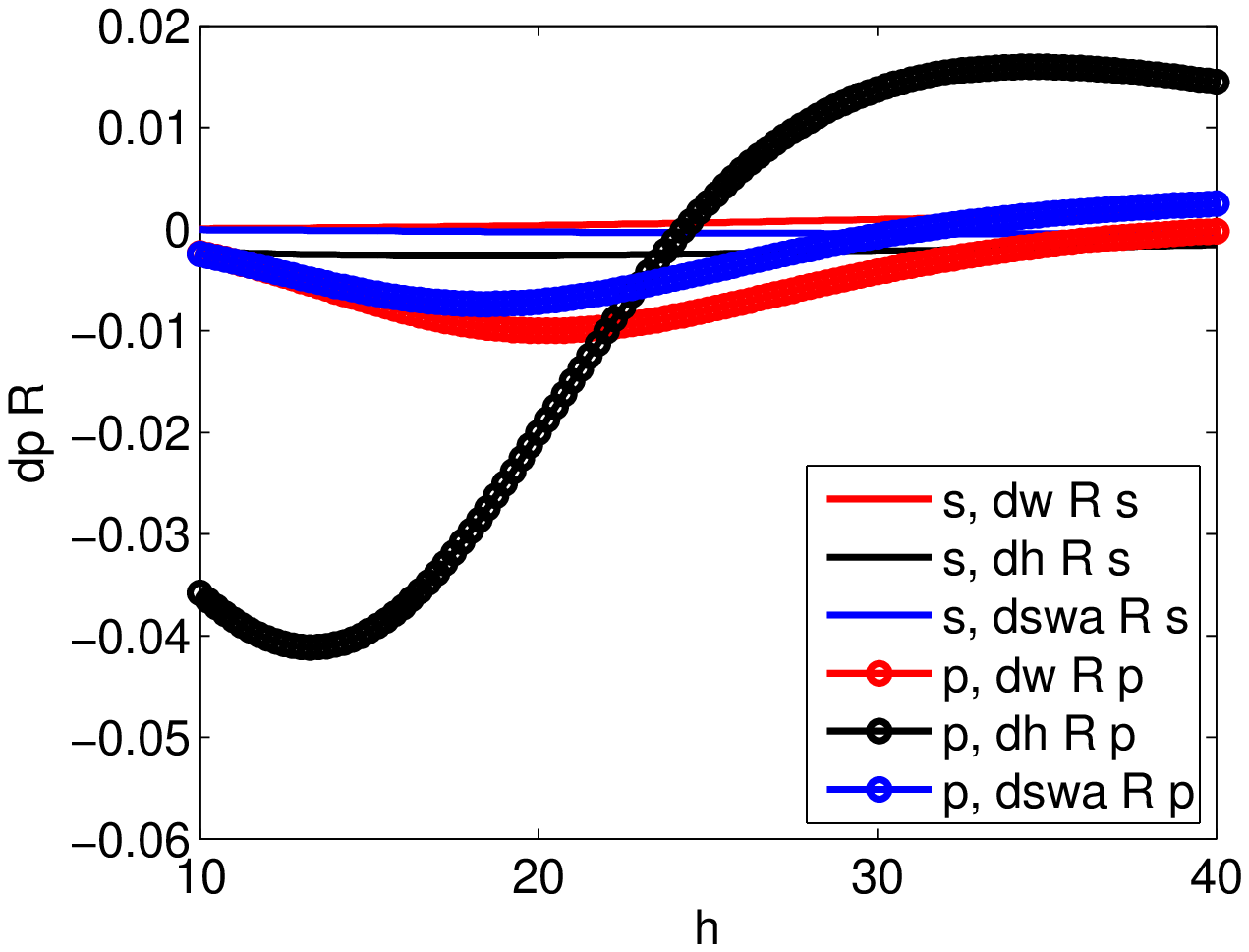}
\end{minipage}
  \caption{
{\it Left:} Dependence of the scatterometric signal $I_0$ on the height of the grating lines $h$, for S- and P-polarization. 
{\it Right:} Dependence of the sensitivity with respect to parameter variations (CD, height, sidewall angle) on $h$. 
}
\label{figure_sensitivity_height}
\end{center}
\end{figure}

Figure~\ref{figure_sensitivity_height} (left) shows how the scatterometric signal (zero order reflection intensity) 
varies with height of the grating lines. As in the previous case, S- and P-polarization show a different behavior. 
The right part of this Figure shows how the sensitivity with respect to parameter variations depends on the line height. 
From this Figure it can, e.g., be seen that in this case again, sensitivity is about an order of magnitude higher for incident P-polarized light, 
and that absolute values of sensitivity with respect to CD variations are highest for line of height $h\approx 20\,$nm
(in the investigated parameter regime).

\section{Conclusion}
To summarize, a method for automatic and computational-cost-effective computation of parameter derivatives 
of electromagnetic near fields and derived quantities has been demonstrated. This is useful for design optimization tasks, 
parameter reconstruction, sensitivity analysis, and other applications. 
A convergence study has been performed which demonstrated that very high levels of accuracy can be achieved. 
The method has been applied to investigate sensitivity of a scatterometric setup in different parameter regimes.

\section*{Acknowledgments}
The work presented here is part of the EMRP Joint Research Project IND\,17 {\sc Scatterometry}.
The EMRP is jointly funded by the EMRP participating countries within EURAMET and the European Union.
The authors would further like to acknowledge the support of
European Regional Development Fund (EFRE) / Investitionsbank Berlin (IBB) through contracts 
ProFIT 10144\,554/5 and the support of DFG (Deutsche Forschungsgemeinschaft) through
the DFG Research Center {\sc Matheon}.

\bibliography{/home/numerik/bzfburge/texte/biblios/phcbibli,/home/numerik/bzfburge/texte/biblios/group,/home/numerik/bzfburge/texte/biblios/lithography,/home/numerik/bzfburge/texte/biblios/group_2012}
\bibliographystyle{spiebib}

\end{document}